\documentclass[12pt]{article}
\usepackage{epsf}
\usepackage{cite}
\usepackage{subeqnarray}

\newcommand{\rf}[1]{(\ref{#1})}
\newcommand{\bea}{\begin{eqnarray}}
\newcommand{\eea}{\end{eqnarray}}
\newcommand{\e}{{\rm e}}

\renewcommand{\b}{\beta}

\newcommand{\del}{\delta}
\newcommand{\D}{\Delta}
\renewcommand{\L}{\Lambda}

\newcommand{\vph}{\varphi}

\newcommand{\oq}{\frac{1}{4}}

\newcommand{\ra}{\right\rangle}
\newcommand{\la}{\left\langle}

\newcommand{\cD}{{\cal D}}

\def\void{}
\def\labelmark{}

\newenvironment{formula}[1]{\def\labelname{#1}
\ifx\void\labelname\def\junk{\begin{displaymath}}
\else\def\junk{\begin{equation}\label{\labelname}}\fi\junk}%
{\ifx\void\labelname\def\junk{\end{displaymath}}
\else\def\junk{\end{equation}}\fi\junk\labelmark\def\labelname{}}

{\ifx\void\labelname\def\junk{\end{array}\end{displaymath}}
\else\def\junk{\end{array}\right.\end{equation}}
\fi\junk\labelmark\def\labelname{}\def\junk{}
}

\newcommand{\beq}{\begin{formula}}
\newcommand{\eeq}{\end{formula}}
\newcommand{\beqv}{\begin{formula}{}}
\newcommand{\bes}{\begin{subeqnarray}}
\newcommand{\ees}{\end{subeqnarray}}

\begin{document}
\setlength{\topmargin}{0pt}
\setlength{\oddsidemargin}{5mm}
\setlength{\headheight}{0pt}
\setlength{\headsep}{0pt}
\setlength{\topskip}{9mm}

\begin{flushright}
\hfill    NBI-HE-96-23\\
\hfill    SU-4240-637\\
\hfill    hep-lat/9606012\\
\hfill    June 96\\
\end{flushright}

\begin{center}
\vspace{24pt}
{\large \bf Geometrical Interpretation of the KPZ Exponents }

\vspace{24pt}

{\sl J. Ambj\o rn } and {\sl K. N. Anagnostopoulos}

\vspace{6pt}

 The Niels Bohr Institute\\
Blegdamsvej 17, DK-2100 Copenhagen \O , Denmark\\

\vspace{12pt}

{\sl U. Magnea} \\
Nordita,\\
 Blegdamsvej 17, DK-2100 Copenhagen \O , Denmark\\

\vspace{6pt}

and

\vspace{6pt}

{\sl G. Thorleifsson} \\
Physics Department, Syracuse University,\\
 Syracuse, NY 13244, USA

\end{center}
\vspace{24pt}

\vfill

\begin{center}
{\bf Abstract}
\end{center}

\vspace{12pt}

\noindent
We provide evidence that the KPZ exponents in two-dimensional
quantum gravity can be interpreted as scaling exponents
of correlation functions which are functions of the invariant geodesic
distance between the fields.

\vfill

\newpage

\section{Introduction}

The calculation of the dressed scaling exponents
is a milestone in the theory of two-dimensional quantum gravity
\cite{kpz,ddk}. Strictly speaking, however,  the derivation uses only
finite size scaling arguments and involves only matter field
correlators integrated over all space--time. The result can be
formulated as follows: if we are given a conformal field theory where
a scalar operator $\phi$ has scaling dimension $\D_0$ we know that the
integrated correlator will scale with the volume $V$ as follows
\beq{*1x}
\int_V d^2 x \int_V d^2 y\; \la \phi (x)\phi (y)\ra
\sim V^{2-2\D_0}.
\eeq
If the conformal field theory is coupled to quantum gravity the
corresponding expectation value now includes the average over
equivalence classes of metrics with $\int d^2 x \sqrt{g(x)} = V$ and
Eq.~\rf{*1x} is replaced by
\beq{*2x}
\la \int_V d^2 x \sqrt{g(x)}\int_V d^2y \sqrt{g(y)}\; \phi(x)\phi (y) \ra
\sim V^{2-2\D},
\eeq
where the {\it dressed} scaling exponent $\D$ for a conformal field
theory with central charge $c$ is related to the scaling exponent
$\D_0$ in flat space by
\beq{*3x}
\D/2 = \frac{\sqrt{1-c + 24 \D_0/2} - \sqrt{1-c}}{\sqrt{25-c} -
\sqrt{1-c}}.
\eeq

While these results are beautiful, one key ingredient is missing
compared to the theory in flat space: the concept of a correlation
length which diverges at the critical point.  In fact, it has
oocasionally been argued that there exists no such concept in quantum
gravity since we integrate over all metrics.  As we discuss below,
however, a proper definition of invariant correlation length exists,
but even with this definition at hand there are situations where the
correlation length is not divergent in quantum gravity, although it
diverges for the same matter system in flat space. For many Ising
spins coupled to quantum gravity there exist convincing arguments
which show that the spin--spin correlation length does not diverge at
the critical point even if the transition is not first order
\cite{ah}.

The task of this article is to provide evidence that there exists a
divergent correlation length associated to correlation functions of
matter fields in two-dimensional quantum gravity coupled to matter
with central charge in the interval $0< c < 1$.  First a notation of
{\it reparametrization invariant distance} is needed in order that one
can attribute a physical meaning to a divergent correlation
length. One trivial suggestion is the following: for a fixed metric
the concept of geodesic distance serves well as the definition of
invariant distance. We can transport this notation to quantum gravity
provided we perform the quantum average over metrics where the two
marked points {\it are} precisely separated by a geodesic distance
$R$. In this way a possible definition of a two--point correlator of
$\phi(x)$ could be
\beq{*1}
G_{\phi} (R;\L) = \int \cD [g] \cD \vph \;\e^{-S_G-S_M}
\int d^2x d^2y \sqrt{g(x)g(y)} \;
\phi(x) \phi(y) \del( D_g(x,y) -R).
\eeq
In Eq.~\rf{*1} $D_g(x,y)$ denotes the geodesic distance between $x$
and $y$ with respect to the given metric $g$.  The integration $\cD
[g]$ is intended to be performed over equivalence classes of metrics,
while $\cD \vph$ signifies the integration over the matter fields
which are coupled to quantum gravity via the action $S_M(\vph)$. $\L$
is the cosmological constant.

While this definition is straightforward it has not been easy to use
in a continuum context, i.e.~in Liouville theory.  An inconclusive
attempt was pioneered by F.~David \cite{david1}, but 
the $\del$-function constraint made it difficult to perform any
detailed calculations.  In the context of two-dimensional gravity and
non-critical string theory, regularized by the use of dynamical
triangulations \cite{adf,adfo,david2,kkm}, it is possible to implement
the $\del$-function constraint in Eq.~\rf{*1}. This was done in
\cite{aw}, using the so called transfer matrix formalism of dynamical
triangulated surfaces \cite{transfer,watabiki,noboru} (see also
\cite{kawai,japanese} for continuum transcriptions of this formalism).
In \cite{aw} the discretized version of Eq.~\rf{*1} was calculated in the
case $\phi= 1$ and in the scaling limit the following result was
obtained:
\beq{*2}
G_{1} (R;\L) = \L^{4/3} \frac{\cosh \L^{\oq} R}{ \sinh^3 \L^{\oq} R},
\eeq
In \cite{ope} this result was generalized to other scaling
operators in pure two-dimensional quantum gravity
and the concept of an operator product expansion was developed, the
distance of separation between the various operators being
geodesic in the same way as in Eq.~\rf{*1}.

As shown in \cite{adj,aw} an object like $G_{1}(R;\L)$ is a
good probe of the fractal structure of space{--}time which is
determined by the exponential decay of $G_1(R;\L)$. We expect
\beq{*2a}
G_{1}(R;\L) \sim e^{-c \L^{1/d_H} R},
\eeq
where $d_H$ is the Hausdorff dimension of our universes in the
following sense: consider the ensemble of universes where two
marked points are separated a geodesic distance $R$. We define
$d_H$ by
\beq{*2b}
\la V \ra_R   \sim R^{d_H}
\eeq
provided $R$ is not much larger than $1/\L^{1/d_H}$.
We conclude from Eq.~\rf{*2} that the fractal dimension of space--time
in two-dimensional quantum gravity is 4, as first proved
by different means in \cite{transfer}.

By a Laplace transformation it is possible to obtain the correlation
functions for universes of fixed volume $V$:
\beq{*3}
G_{\phi}(R;\L) = \int_0^\infty dV \; \e^{-\L V} G_{\phi}
(R;V).
\eeq
For $\phi=1$ the correlation function $G_{1} (R;V)$ is related to the
average ``area'' of spherical shells $ S_1(R;V)$ of geodesic radius
$R$ by
\beq{*4}
S_1(R;V) = \frac{1}{V}\, \frac{G_{1} (R;V)}{Z(V)},
\eeq
where $Z(V)$ denotes the partition function of fixed volume, i.e.
\beq{*5}
Z(V) = \int \cD [g] \cD \vph \; \e^{-S_G-S_M} \del(\int d^2x \sqrt{g(x)} -V).
\eeq
An alternative and very natural definition of Hausdorff dimension is
\beq{*5a}
S_1(R;V)  \sim R^{d_h-1}~~~~{\rm for}~~~R \ll V^{1/d_h}.
\eeq
By an inverse Laplace transformation of Eq.~\rf{*2} it follows that
$d_h =d_H=4$.  However, in the case of matter fields coupled to
gravity there exists no proof that $d_H = d_h$. If $d_H=d_h$ it is
natural to introduce the finite size scaling variable
\beq{*6}
x = \frac{R}{V^{1/d_h}}.
\eeq
From Eqs.~\rf{*5a} and \rf{*2a} we expect for $S_1(R;V)$ a finite size
scaling relation of the form
\beq{*7}
S_1(R;V) \sim V^{1-1/d_h} \; F(x) \; ,
\eeq
where $F(x)$ has the following properties:
\beq{*8}
F(x) \sim x^{d_h-1} ~~~{\rm for}~~~x \ll 1,~~~~~~~~~F(x) \to 0 ~~~{\rm for}
~~~x \gg 1.
\eeq
These relations are of course satisfied with $d_h=4$ in pure gravity
\cite{aw,ajw} and numerical simulations indicate that it is also true
for $0 < c < 1$ \cite{syracuse,ajw}. The finite size scaling arguments
can also be applied to the more general correlators
\beq{*9}
S_\phi(R;V) \equiv  \frac{G_{\phi} (R;V) }{V \, Z(V)}.
\eeq
From Eq.~\rf{*3x} we have for a field with scaling exponents $\D_0$ in
flat space and $\D$ after coupling to quantum gravity \cite{syracuse,ajw}:
\beq{*10}
\int_0^\infty dR \; S_\phi(R;V) \sim V^{1-\D}.
\eeq
From this scaling and the definition of $d_h$
it is tempting to conjecture the following scaling behavior:
\beq{*11}
S_\phi(R; V) \sim \frac{R^{d_h-1}}{ R^{d_h \D}} f(x) =
V^{1-\D-1/d_h} F_\phi(x),
\eeq
where $f(0) >0$ and $F(x) \sim x^{d_h(1-\D) -1}$ for small $x$.
This formula generalizes Eqs.~\rf{*5a}-\rf{*8} to the case where matter is
coupled to quantum gravity. The corresponding formula
in flat two-dimensional space reads:
\beq{*12}
S^{(0)}_\phi(R;V) \sim  \frac{R}{R^{2\D_0}} f(R/\sqrt{V}).
\eeq

In this article we report on extensive computer simulations
where we tried to verify Eq.~\rf{*11}. The finite size aspect
of the formula, i.e. the volume dependence, has already been
checked in \cite{syracuse}. However, until now no reliable results
exist for the correlator itself, i.e. for the $R$ dependence.
As explained above this $R$ dependence, i.e. the infinite
correlation length,  is the crucial feature of conformal field
theory and we provide evidence that it exists also for
two-dimensional quantum gravity coupled to a conformal field
theory if we use the geodesic distance as a measure of length.

\section{Numerical setup and results}

The numerical simulations are performed as follows: As discretization
we use dynamical triangulations. In this formalism surfaces with
spherical topology are constructed from equilateral triangles and we
allow self energy and tadpole graphs to form in the dual $\varphi^3$
graph of the triangulation.  To each {\it vertex} is associated a
spin, either Ising or three-states Potts spin, depending on the
model. A given spin interacts with the spins on the neighboring
vertices. The Monte Carlo updating of the triangulations is performed
by the so{--}called flip algorithm and the spins are updated by
standard cluster algorithms. The flips are organized in ``sweeps''
which consist of approximately $N_L$ {\it accepted} flips where $N_L$
is the number of links of the triangulated surface.  After a sweep we
update the spin system.  All this is by now standard and we refer to
\cite{bj,ckr,adjt,atw} for details about the actions or Monte Carlo
procedures.

The results presented in this paper cover system sizes from 16000 to
128000 triangles and the number of sweeps is $1.7${--}$5.0\times
10^6$.  For the Ising and three-states Potts model on dynamically
triangulated surfaces the critical temperatures $\b_c$ are known
\cite{bk,daul}. We use these values in the simulations.

Geodesic distances on the triangulations are defined either as the
shortest link distance between two vertices or the shortest path
through neighboring triangles. While these two distances can vary a
lot for individually chosen points on a given triangulation, they are
proportional when the average over the ensemble of triangulations is
taken. We will report here the results obtained by the use of link
distances. Our discretized length $r$ will be the number of links. Our
discretized volume (or area) will be the number $N_T$ of triangles
used in the computer simulations and our discretized scaling variable
will be
\beq{*20}
x \equiv \frac{r}{N^{1/d_h}}\, ,
\eeq
where $N$ is the number of vertices on the surface.
We denote the discretized distributions corresponding to $S_{1}(R;V)$
and $S_{\phi}(R;V)$ by
\bes
\label{*21all}
\slabel{*21}
n_1    (r;N) &=& \langle \sum_j \,\delta(D_{ij}-r) \rangle\, ,\\
\slabel{*21a}
n_\phi (r;N) &=& \langle \sum_j \,\sigma_i\sigma_j\,\delta(D_{ij}-r) \rangle
   \, ,
\ees
where the indices $i$ and $j$ label vertices: $i$ is a random fixed
vertex for each measurement and $j$ runs over all vertices of the
given configuration.  $D_{ij}$ is the link distance between the
vertices labelled by $i$ and $j$.

Our first task is to determine $d_h$. If the scaling hypothesis is
correct, both the distributions $n_{1} (r;N)$ and $n_\phi (r;N)$ will
depend only on the scaling variable $x$ given by Eq.~\rf{*20}, except for
the overall scaling factors $N^{1-1/d_h}$ and $N^{1-\D -1/d_h}$
respectively. Hence, we have to determine the best value of $d_h$ such
that 
\beq{*22} n_1 (r;N) = N^{1-1/d_h} F(x),~~~~~~n_\phi (r;N) =
N^{1-\D-1/d_h} F_\phi(x) 
\eeq 
for all values of $r$ and $N$.  This has already been done for the
Ising model and the three-states Potts model for the $n_1(r;N)$
correlator in \cite{syracuse,ajw} and for $n_\phi(r;N)$ in
\cite{syracuse}.  Our data is in very good agreement with the results
obtained in \cite{syracuse,ajw}, namely $d_h=4$ for both the Ising
model and the three-states Potts model coupled to gravity. Details of
these measurements will be published elsewhere \cite{preparation}.
\begin{figure}[ht]
\centerline{\epsfxsize=4.0in \epsfysize=2.67in \epsfbox{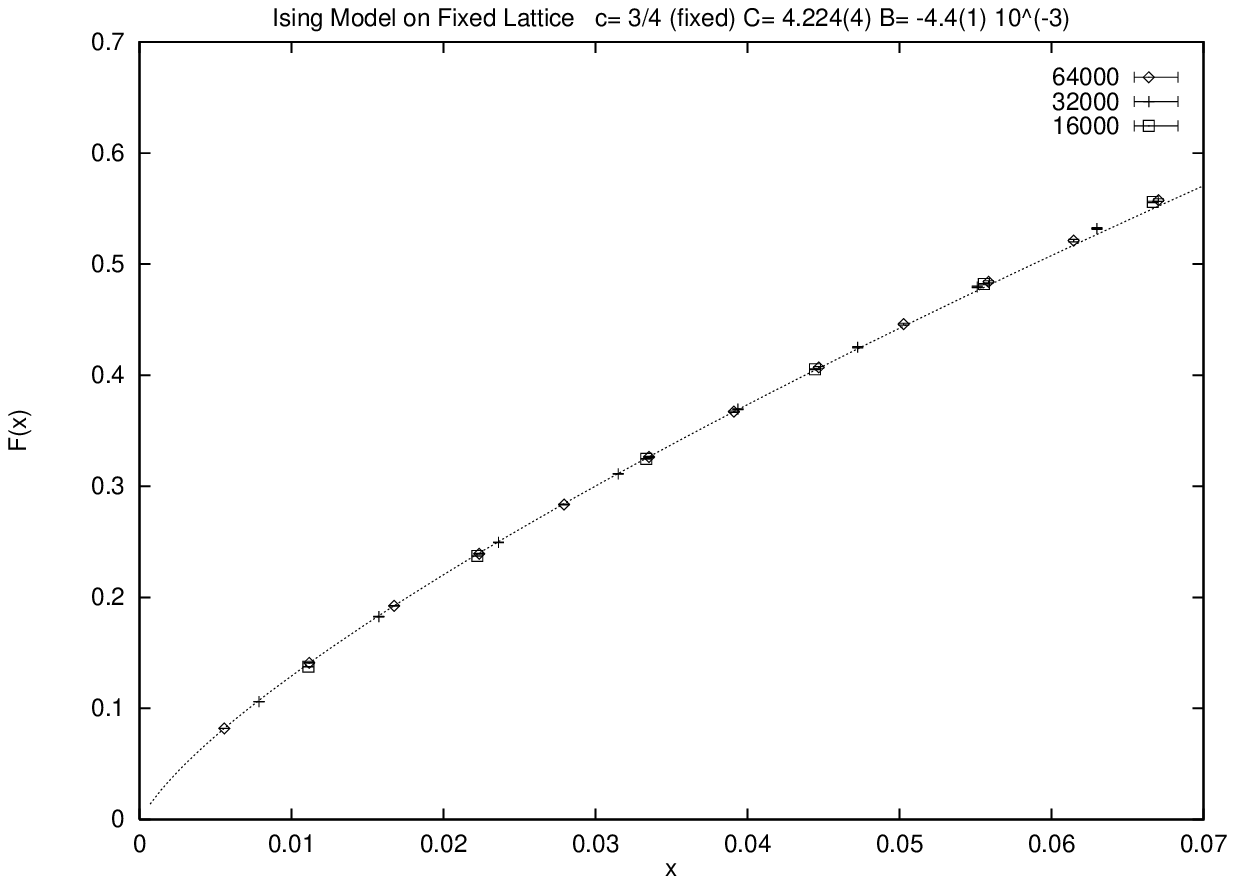}}
\centerline{\epsfxsize=4.0in \epsfysize=2.67in \epsfbox{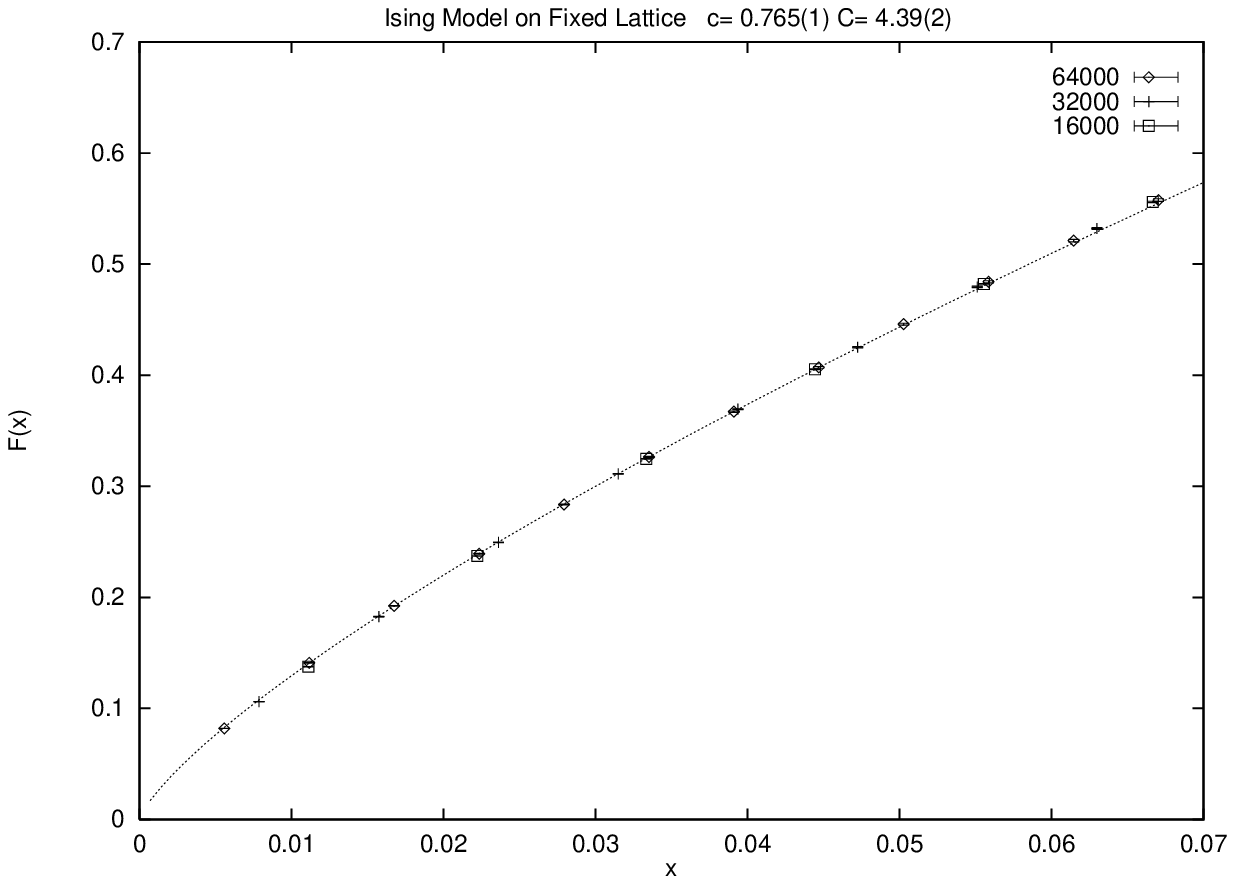} }
\caption{ ({\it a}) Data for
$F^{Flat}_\phi(x)$, as 
defined in Eq.~\rf{*26}, for small values of $x=\frac{r}{N^{1/2}}$ in
the case of the Ising model on a fixed triangular
lattice with $T^2$ topology. The fit shown is to Eq.~\rf{*27} for
$N_T=64000$ triangles. ({\it b}) Same as ({\it a}) but the fit is 
to Eq.~\rf{*27a}.  }
\label{f:1}
\end{figure}
\begin{figure}[ht]
\centerline{\epsfxsize=4.0in \epsfysize=2.67in \epsfbox{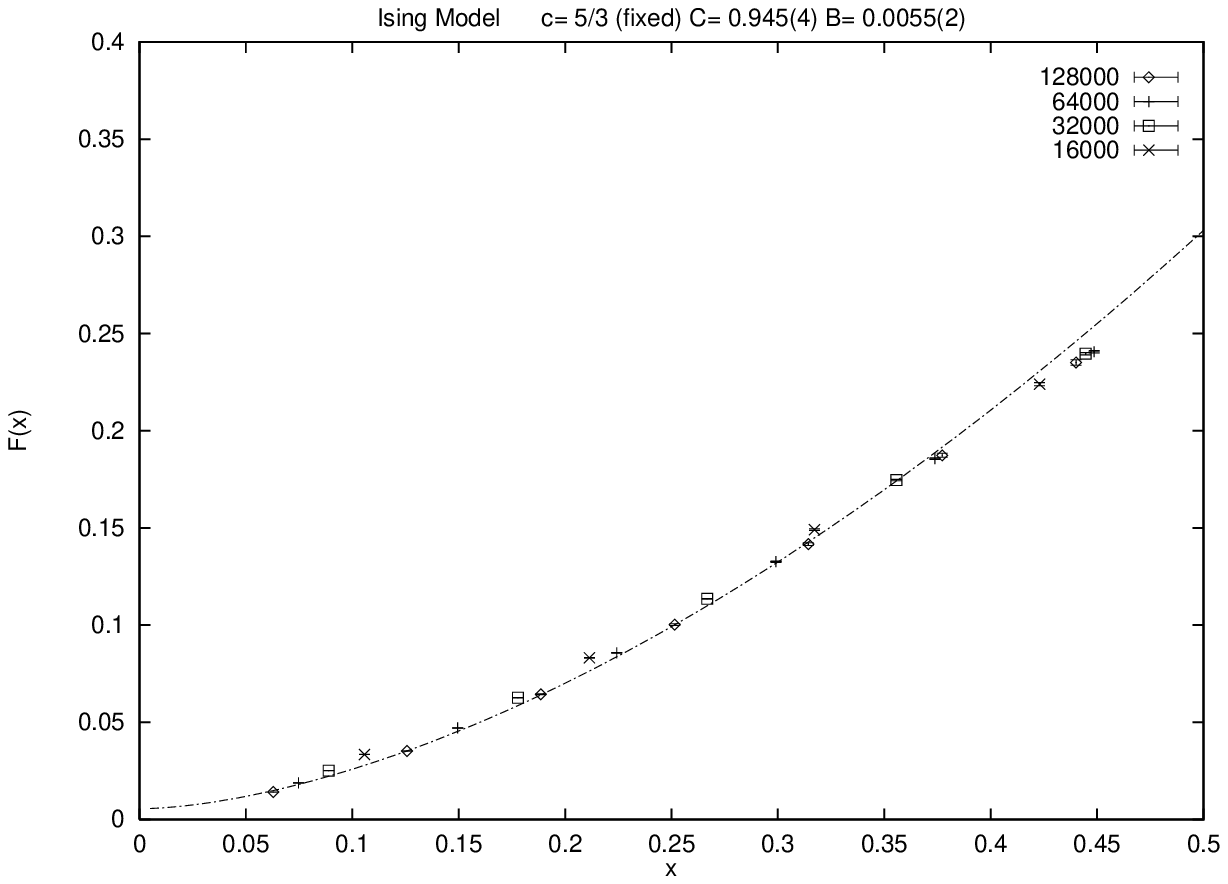}}
\centerline{\epsfxsize=4.0in \epsfysize=2.67in \epsfbox{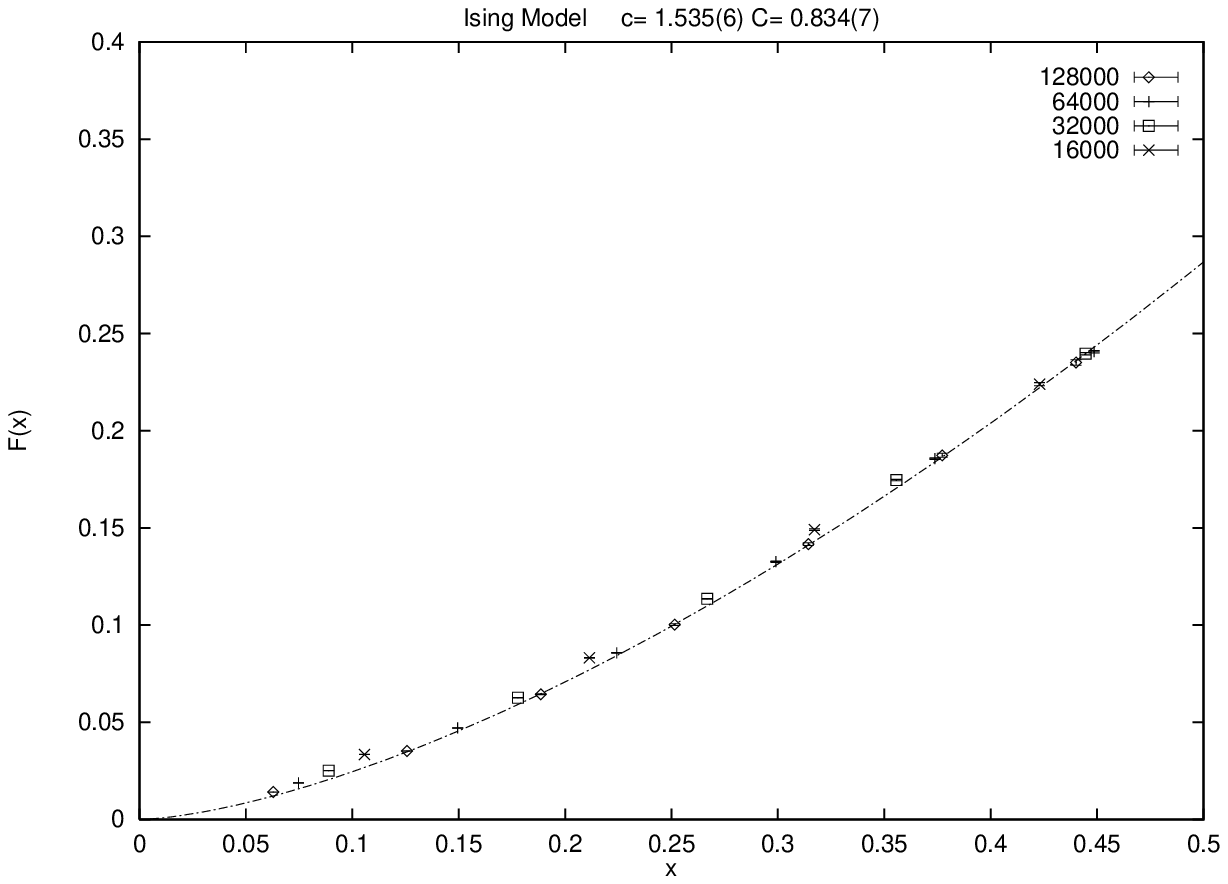} }
\caption{({\it a}) Same as Fig.~\ref{f:1} but for the Ising model 
coupled to quantum gravity. The fit shown is to Eq.~\rf{*27} with
exponent $4/3$ for $N_T=128000$ triangles. 
({\it b}) Same as ({\it a}) but the fit is now to Eq.~\rf{*27a}. }
\label{f:2}
\end{figure}
\begin{figure}[ht]
\centerline{\epsfxsize=4.0in \epsfysize=2.67in \epsfbox{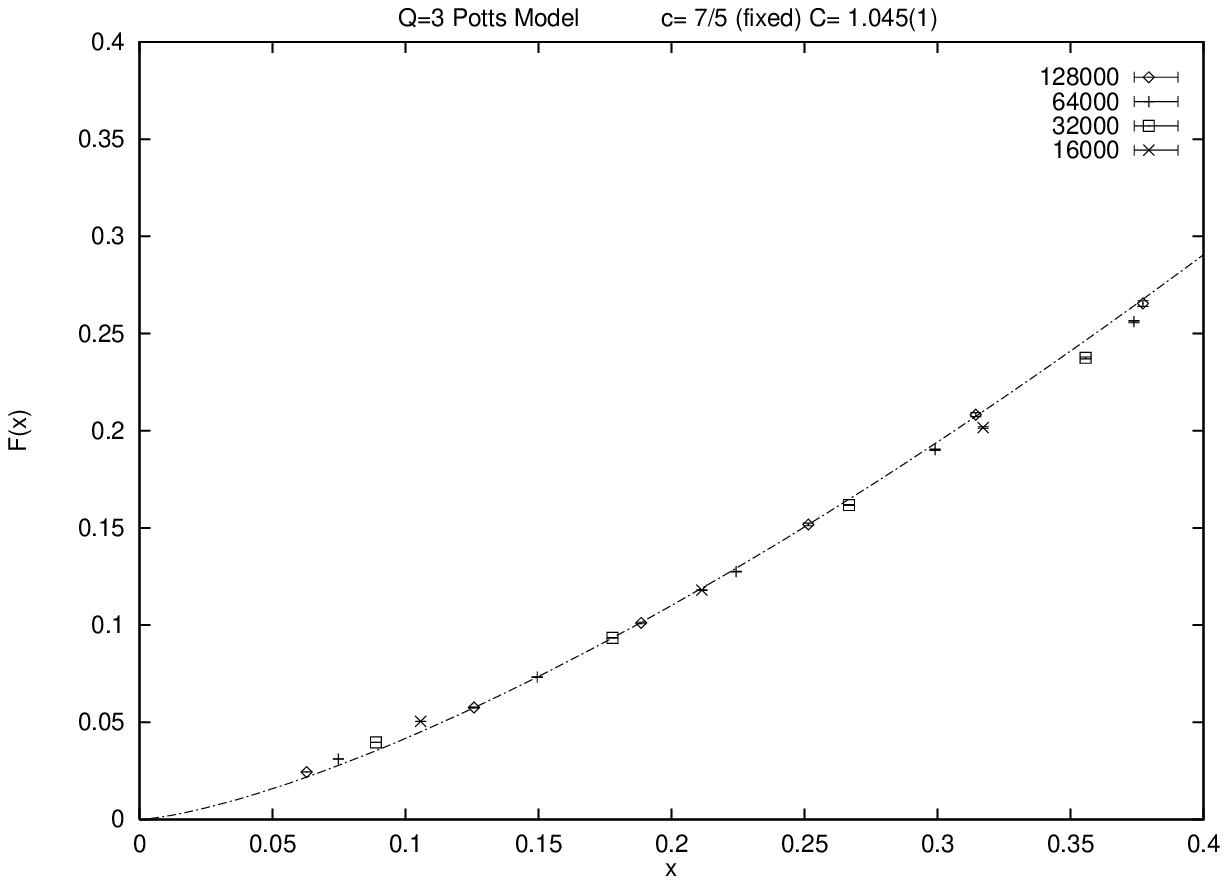}}
\centerline{\epsfxsize=4.0in \epsfysize=2.67in \epsfbox{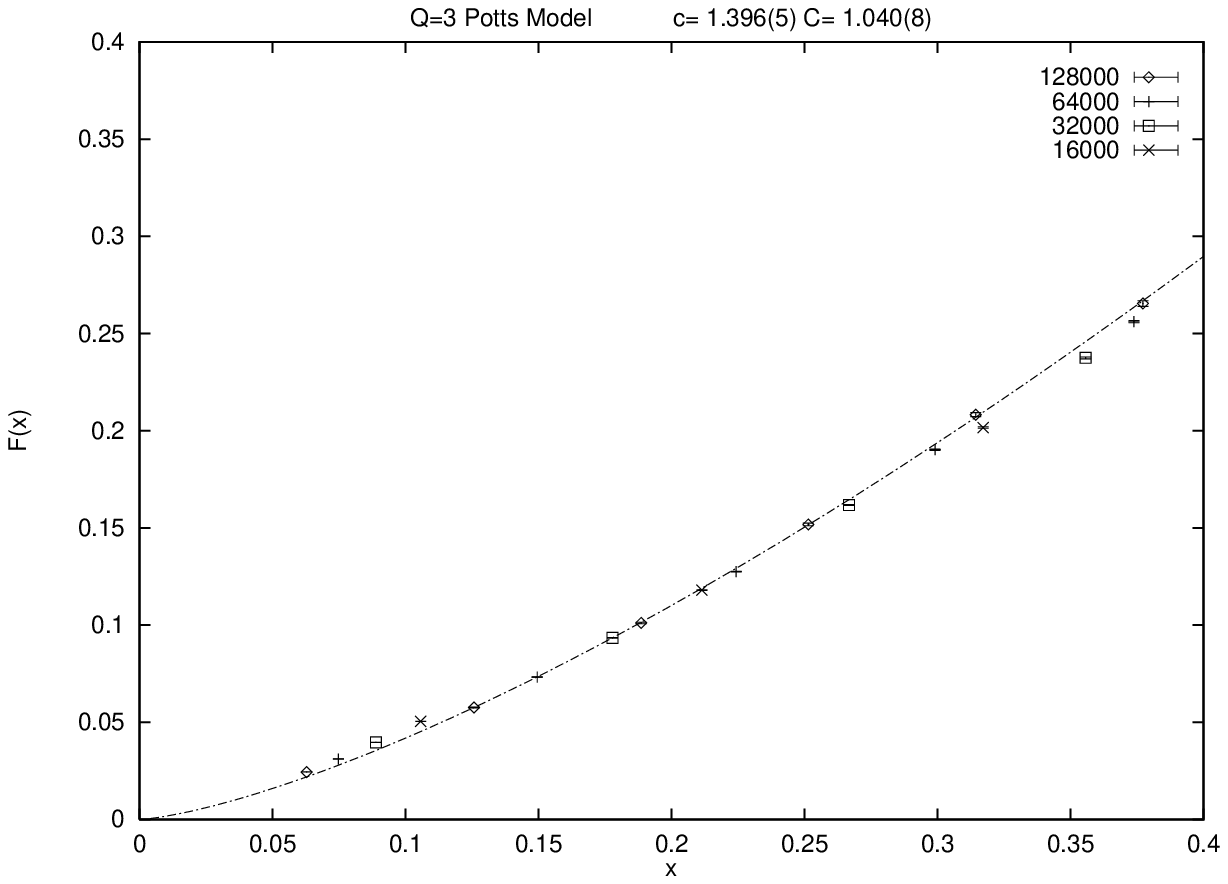} }
\caption{({\it a}) Same as Fig.~\ref{f:1}a in the case of 
the three-states Potts model 
coupled to quantum gravity. The fit shown is to Eq.~\rf{*27} with
exponent $7/5$ for $N_T=128000$ triangles. 
({\it b}) Same as in ({\it a}) but the fit is to Eq.~\rf{*27a}.}
\label{f:3}
\end{figure}

Let us now turn to the main task, the determination of the short
distance behavior of the spin-spin correlator for the Ising model and
the three-states Potts model coupled to gravity. The Ising model on a
regular lattice has a second order transition at a critical
temperature $\b_c^{flat}$ and the corresponding conformal field
theory has central charge $c=1/2$. The spin field has scaling
dimension $\D_0 = 1/8$ and the dressed scaling dimension after
coupling to two-dimensional quantum gravity is $\D= 1/3$. The
tree-states Potts model corresponds to a $c= 4/5$ conformal field
theory and the dressed scaling dimension of the spin field in this
case is $\D=2/5$.  The prediction for short distance scaling is then
\beq{*23}
F_\phi (x)  ~\propto~   x^{d_h(1-\D) -1} ~~~~~{\rm for} ~~~x \ll 1,
\eeq
i.e. (using the value $d_h=4$)
\bes
\label{*24all}
\slabel{*24}
F_\phi^{Ising}(x)  &\propto & x^{5/3}~~~~{\rm for}~~~x \ll 1 \\
\slabel{*25}
F_\phi^{Potts}(x)  &\propto & x^{7/5}~~~~{\rm for}~~~x \ll 1\, .
\ees

In order to calibrate the expected accuracy with which the
exponent $d_h(1-\D)-1$ can be extracted  we have
performed simulations of an Ising spin system on a flat lattice
(with periodic boundary conditions) and measured the known function
\beq{*26}
F_\phi^{Flat} (x) ~\propto~ x^{3/4}~~~~{\rm for}~~~x \ll 1,
\eeq
where $x = r/\sqrt{N}$ and $r$ is the distance in lattice units
between a spin and its ``spherical shell''. This rather unusual way of
measuring correlation functions works quite well as one can see from
the results shown in Table\ \ref{t:1}.  We performed simulations on
regular toroidal triangular lattices of sizes $16000${--}$64000$
triangles. In order to understand the size of finite size effects and
the effect of the function $f(x)$ we show the results of the fits to
the following functional forms:
\bes
\label{*27all}
\slabel{*27}
F^{Flat}_\phi (x) &=&  C x^{3/4} +B \, ,\\
\slabel{*27a}
                  &=&  C x^{c} \, ,\\
\slabel{*27b}
                  &=&  C x^{c} + B\, .
\ees
The constant $B$ in Eqs.~\rf{*27} and \rf{*27b} reflects the fact that
the volume element is discrete with a smallest unit and the spin
correlated with itself in the volume element at the shortest distance
is 1.  The data is shown in Fig.~\ref{f:1} together with the best fit
to the largest lattice. It is clear that the determination of the
exponent $c$ is in excellent agreement with the theory. It is
important that we only use $x \ll 1$ if we fit to Eqs.~\rf{*27all}. It
is pleasant surprise that the continuum formulae are valid even for
small values of $r$.  For $x > 0.06$ the function $f$ present in
Eq.~\rf{*12} will play an important role.

For the Ising model and the three-states Potts model coupled to
gravity we repeat the analysis performed for the regular lattice. We
fit the data to Eqs.~\rf{*27}, with exponents $5/3$ and $7/5$
respectively, \rf{*27a} and \rf{*27b} for $x < 0.45$.  The results are
shown in Table~\ref{t:2} for the Ising model and Table~\ref{t:3} for
the three{--}states Potts model. The corresponding plots including the
best fit to the largest lattice are shown in
Fig.~\ref{f:2}{\it a} and Fig.~\ref{f:2}{\it b} for the Ising model
and in Fig.~\ref{f:3}{\it a} and Fig.~\ref{f:3}{\it b} for the
three-states Potts model.

Our results are consistent with the exponents conjectured in
Eqs.~\rf{*24} and \rf{*25}.  We find that simulating the largest
lattices, 128000 triangles, is important for obtaining enough data
points in the relevant region $x < 0.45$.  The fits to the predicted
behavior Eq.~\rf{*27} are good and the finite size correction $B$
approaches zero as the volume increases.  Even if the exponent $c$ is
allowed to vary, as in the fits to Eqs.~\rf{*27a} and \rf{*27b}, it
approaches its predicted value convincingly as the volume is
increased. The small discrepancy is consistent with finite size
effects, which clearly are more important than the statistical errors
quoted in Tables~\ref{t:2} and \ref{t:3}. We find that the difference
in the values of $c$ obtained from the fits to Eqs.~\rf{*27a} and
\rf{*27b} gives a measure of the systematic errors entering from
varying the range and type of the fits.

\begin{table}[ht]
\begin{center}
\begin{tabular}{|c|c c| c c |c c c|}\hline
$N_T$  &  $C$    &  $B$      & $c$       & $C$     &  $c$   &  $C$  &  $B$ \\ 
\hline\hline
16000& 4.276(5)& -0.0088(2)& 0.7806(8)& 4.62(2)& 0.754(7)& 4.32(7)& -0.008(1)\\ 
32000& 4.248(4)& -0.0062(2)& 0.769(1) & 4.44(2)& 0.756(4)& 4.32(5)& -0.005(1)\\ 
64000& 4.224(4)& -0.0044(1)& 0.765(1) & 4.39(2)& 0.753(4)& 4.25(4)& -0.0038(7)\\
\hline
\end{tabular}
\end{center}
\caption{ Results for the fits to Eqs.~\rf{*27}, \rf{*27a} and 
\rf{*27b} for the Ising model on a fixed triangular lattice with $T^2$
topology.}
\label{t:1}
\end{table}
\begin{table}[ht]
\begin{center}
\begin{tabular}{|c|c c|c c|c c c|}\hline
$N_T$   & $C$     & $B$      & $c$     & $C$     &  $c$    &  $C$    &  $B$ \\
\hline\hline
16000 & 0.880(4)& 0.0173(4)& 1.432(6)& 0.770(6)& 1.51(1) & 0.799(8)& 0.0063(4)\\
32000 & 0.923(3)& 0.0108(2)& 1.470(3)& 0.793(3)& 1.545(5)& 0.829(4)& 0.0054(1)\\
64000 & 0.940(3)& 0.0075(1)& 1.497(4)& 0.805(4)& 1.565(4)& 0.841(4)& 0.0043(1)\\
128000& 0.945(4)& 0.0054(2)& 1.535(6)& 0.834(7)& 1.596(6)& 0.870(7)& 0.0036(1)\\
\hline
\end{tabular}
\end{center}
\caption{Same as in Table~\ref{t:1}
for the Ising model coupled to quantum gravity.}
\label{t:2}
\end{table}
\begin{table}[ht]
\begin{center}
\begin{tabular}{|c|c c|c c|c c c|}\hline
$N_T$   & $C$     & $B$      & $c$     & $C$     & $c$     &  $C$    &  $B$ \\
\hline\hline
16000 & 0.947(1)& 0.0098(1)& 1.291(4)& 0.880(5)& 1.331(8)& 0.893(6)& 0.0056(6)\\
32000 & 0.993(3)& 0.0050(3)& 1.348(3)& 0.959(5)& 1.425(7)& 1.007(7)& 0.0077(3)\\
64000 & 1.013(2)& 0.0024(2)& 1.371(2)& 0.991(4)& 1.447(5)& 1.047(6)& 0.0065(2)\\
128000& 1.044(4)& 0.0003(3)& 1.396(5)& 1.040(8)& 1.470(8)& 1.10(1) & 0.0055(3)\\
\hline
\end{tabular}
\end{center}
\caption{Same as in Table~\ref{t:1}
for the three-states Potts model coupled to quantum gravity.}
\label{t:3}
\end{table}

\section{Discussion}

We have provided substantial evidence for the conjecture put forward
in \cite{ajw} that the the two-point correlator in a unitary conformal
field theory translates the two-point correlator of the same theory
coupled to quantum gravity in the following way:
\beq{*30}
S_\phi^{flat}(R;V) =  \frac{R}{R^{2\D_0}} f^{flat}(R/\sqrt{V})
~~\to~~
S_\phi(R;V) = \frac{R^{d_h-1}}{R^{d_h \D}} f (R/V^{1/d_h}),
\eeq
where $f(0) >0$, $R$ is the geodesic distance as defined above and
$d_h$ is the fractal dimension of space-time.

A number of questions still need to be answered before we have a
complete understanding of the concept of invariant correlation
functions in two-dimensional quantum gravity.  First, the value of
$d_h$ as a function of the central charge $c$ of the conformal field
theory should be clarified.  Some formal arguments suggest that
$d_h=2m$ for a $(m,m+1)$ conformal theory coupled to gravity. This
would imply that $d_h \to \infty$ for $c = 1-6/m(m+1) \to 1$. For the
three-states Potts model $m=4$, i.e. the value of $d_h=8$.  This is
definitely not seen in the computer simulations, where particularly
the data for $n_1 (r;N)$ favor $d_h=4$. However, one could argue that
the systems considered so far are much too small for observe $d_h=8$,
since $(128000)^{1/8} \approx 4$. While this might be true, it
contradicts the fact that finite size scaling works very well for the
three-states Potts model coupled to quantum gravity already for
considerable smaller systems and we get the correct critical
exponents. All the scaling arguments of \cite{syracuse,ajw} yield
$d_h=4$ with quite small error; the only significant discrepancy being
that of the height of the maximum of the distribution $n_\phi(r;N)$
for the three-states Potts model. According to Eq.~\rf{*22} it should
scale as $N^{d_h(1-\D)-1}$, but using the theoretical value $\D = 7/5$ we
obtain $d_h = 4.32(2)$ from collapsing the $64K${--}$128K$
configurations. Finite size effects, however, play an important role
since the value of $d_h$ extracted this way decreases rapidly with
system size towards the value $4$. We will report more details in the
future \cite{preparation}.

Next, one should understand the tail of the distribution $n_\phi
(r;N)$ for $r \gg N^{1/d_h}$. The following simple argument indicates
that the distribution should be identical to that of pure gravity: for
such large values of $R$ the universes we observe are essentially
one-dimensional, since they have to be long tubes of almost no
transverse extension. One-dimensional universes cannot (contrary to
genuine two-dimensional universes) have non-trivial matter
interactions unless they are long range, which is not the case
here. Consequently there should be no difference between pure gravity
and gravity including matter fields.  This is only a heuristic
argument, but if true it rules out the existence of a single scaling
variable $x$ as in Eq.~\rf{*20} for the whole range of $x$ unless $d_h=4$,
which is the value for pure gravity.  The argument might be too simple
since preliminary results for $c=-2$ conformal field theories coupled
to gravity indicate both a common scale variable for all values of $x$
and a $d_h < 4$ \cite{progress}. The $c=-2$ matter theory is on the
other hand non-unitary and the heuristic arguments might be still be
valid for unitary theories.

Finally, it would be most interesting to be able to prove analytically
the results reported above and extend the concept of operator product
expansion initiated in \cite{ope} to the case of unitary conformal
field theories coupled to two-dimensional quantum gravity.

\end{document}